\documentclass[aps,twocolumn]{revtex4-1}
\usepackage[T1]{fontenc}
\usepackage[latin9]{inputenc}
\usepackage{amsthm}
\usepackage{amsmath}
\usepackage{graphicx}
\usepackage{multirow}

\begin{document}

\title{Deep learning for flow observables in ultrarelativistic heavy-ion collisions}

\author{H.~Hirvonen${}^{a,b}$, K.~J.~Eskola${}^{a,b}$, H.~Niemi${}^{a,b}$,  }
\affiliation{$^{a}$University of Jyv\"askyl\"a, Department of Physics, P.O. Box 35, FI-40014 University of Jyv\"askyl\"a, Finland}
\affiliation{$^b$Helsinki Institute of Physics, P.O.Box 64, FI-00014 University of
Helsinki, Finland}

\begin{abstract}
We train a deep convolutional neural network to predict hydrodynamic results for flow coefficients, average transverse momenta and charged particle multiplicities in ultrarelativistic heavy-ion collisions from the initial energy density profiles. We show that the neural network can be trained accurately enough so that it can reliably predict the hydrodynamic results for the flow coefficients and, remarkably, also their correlations like normalized symmetric cumulants, mixed harmonic cumulants and flow-transverse-momentum correlations. At the same time the required computational time decreases by several orders of magnitude. To demonstrate the advantage of the significantly reduced computation time, we generate 10M initial energy density profiles from which we predict the flow observables using the neural network, which is trained using 5k, and validated using 90k events per collision energy. We then show that increasing the number of collision events from 90k to 10M can have significant effects on certain statistics-expensive flow correlations, which should be taken into account when using these correlators as constraints in the determination of the QCD matter properties.
\end{abstract} 
 
\pacs{25.75.-q, 25.75.Nq, 25.75.Ld, 12.38.Mh, 12.38.Bx, 24.10.Nz, 24.85.+p} 
 
\maketitle

\section{Introduction}

Probing the properties of the strongly interacting matter close to a zero net-baryon density is the primary goal of the highest-energy ultrarelativistic heavy-ion collision experiments. One of the most important tools in interpreting the experimental data is relativistic hydrodynamics. In the hydrodynamic limit the matter behavior is controlled by the matter properties like equation of state and transport coefficients such as shear and bulk viscosity. It has been well established that in heavy-ion collisions flow-like signatures are seen in azimuthal angle spectra of produced particles. This indicates that a small droplet of deconfined phase of QCD-matter called quark gluon plasma (QGP) is created in these collisions, and that it exhibits a fluid-like behaviour~\cite{Heinz:2013th, Gale:2013da, Huovinen:2013wma, Shen:2020gef}. 

Comparing the measurements with the predictions of hydrodynamic computations gives then a possibility to determine the QCD matter properties. A reliable estimate of the QCD matter properties with well-defined error bars demands a global analysis of as many experimental observables and collision systems as possible. In the recent years, such global analyses have given constraints on the QCD transport properties~\cite{Gale:2012rq, Niemi:2015qia, Bernhard:2016tnd, Bass:2017zyn, Bernhard:2019bmu, JETSCAPE:2020mzn, Nijs:2020roc, Auvinen:2020mpc, Parkkila:2021tqq, Parkkila:2021yha}. In particular, the shear viscosity near the QCD transition temperature $T \sim 155$ MeV is rather well constrained. For the full temperature dependence of shear viscosity, and especially bulk viscosity, the uncertainties are significantly larger.

A way to improve the analysis is to consider more observables. One challenge here is that in practice it is necessary to compute the hydrodynamic evolution event-by-event, i.e.\ for each collision event separately, so that the computed observables are obtained as averages over a large number of collisions to closely match with the actual measurements. The non-trivial dependence of the final observables on the equation of state, transport coefficients, initial conditions, and the details of the conversion of the fluid to particles together with numerically demanding hydrodynamic simulations makes the global analysis a very CPU intensive task. In particular, this is the case when the global analysis takes into account observables that require high statistics obtained by accumulating a large number of computed collision events.

The most basic experimental observables quantifying magnitude and details of the flow-like behavior are the Fourier coefficients of a azimuthal hadron spectra, which are usually referred to as flow coefficients $v_n$. They are measured as multiparticle correlations. The increased luminosity in recent measurements, especially at the LHC, has enabled precision measurements of multi-particle correlations between flow coefficients all the way up to eight-particle level. To obtain reliable estimates of these correlations from the fluid dynamical simulation can require gathering statistics from $\sim 10^6 $ collision events. Obtaining such a high statistics is computationally very expensive and performing computations gets even more expensive when sampling the $\mathcal{O}(15)$-dimensional parameter space of a global analysis, where statistics should be obtained for around 300 different parametrizations. Typically one event needs $\sim 30$ min computing time from a CPU and thus the total time it would take to perform high statistic global analysis is around $0.5\times10^6\times300 \sim 10^8 $ CPU hours.

One way to decrease the computation time would be to convert the codes to GPU and use a modern GPU based supercomputer to do the computing. Even though this would significantly speed up the simulations, the task would still require a significant amount of computing time. Another possibility is to simplify the complicated fluid dynamical computations and construct fast estimators that can give good estimates of the final state observables from the initial state alone. The simple version of such an estimator for flow coefficients could be constructed for example by assuming a linear relation between initial state eccentricities and corresponding flow coefficients. As shown in Refs.~\cite{Gardim:2011xv, Niemi:2012aj, Niemi:2015qia} this kind of linear relation works reasonably well for $v_2$ in central collisions, but non-linear effects start to get noticeable in more peripheral collisions and even more so in the case of higher-order flow coefficients for which this kind of estimator would not work well even to begin with.

In this article we present a way to estimate $p_T$-integrated flow observables and correlators directly from the initial energy density profile based on deep convolutional neural networks (CNN). The convolutional neural networks have been proven to be very efficient and accurate tools when it comes to image classification and computer vision tasks. During the past decade, network architectures have evolved towards deeper and deeper networks, i.e.\ a typical network contains more layers than before. A modern CNN architecture can contain hundreds of layers and tens of millions trainable parameters. Neural networks and deep learning has been utilized before in the context of heavy-ion collisions for various different applications, such as impact parameter estimation, identifying quenched jets or determination of the QCD matter phase transition~\cite{David:1994qc, Bass:1996ez, Xiang:2021ssj, Liu:2022hzd, Pang:2016vdc}. In Ref.~\cite{Huang:2018fzn} it was shown that the neural network can also model full hydrodynamic evolution on short time periods, $ \Delta \tau \approx$ 2 fm, but this kind of method has not yet been applicable for modelling a complete space-time evolution of QGP. Deep neural network was also applied to estimating $v_2$ from the kinematic information of particles in the context of the AMPT-model~\cite{Mallick:2022alr}. However, until the current study, neural networks have not been successfully trained to predict flow observables and correlators from the initial state energy density.

The basic setup here is the perturbative QCD based EKRT (Eskola-Kajantie-Ruuskanen-Tuominen) gluon saturation model~\cite{Eskola:1999fc, Paatelainen:2012at} for the computation of initial conditions that, when supplemented by relativistic hydrodynamic evolution~\cite{Paatelainen:2013eea, Niemi:2015qia}, gives a good overall description of the available flow data from heavy-ion collisions at RHIC and LHC~\cite{Niemi:2015voa, Eskola:2017bup, Hirvonen:2022xfv}. The neural network constructed here is, however, not restricted to this particular model, but can in principle be applied to any similar framework.

This paper is organized in the following way: In Sec.~\ref{sec:setup} we briefly go through the structure of the used neural network and give details about how it is implemented in practice. In Sec.~\ref{sec:validation} we validate the accuracy of the neural network by showing that the results obtained by the network match well with the hydrodynamic simulations. The main results are then shown in Sec.~\ref{sec:results}, where we present the neural network predictions for various different correlators with 10M generated collision events. The summary and conclusions are then given in Sec.~\ref{sec:conclusions}.

\section{Model setup}
\label{sec:setup}
\subsection{DenseNet}
Evolution of CNN architectures towards deeper networks has caused challenges to their design~\cite{Alzubaidi21}. 
Very deep networks can easily lose some information about the input. Additionally, when propagating the gradient information from the output back to the input, the gradients can start to approach zero. Therefore, the optimizer leaves the network weights close to the input nearly unchanged so that the loss function won't converge to the global minima. This makes the training of a model slow and inaccurate. To solve the vanishing gradient and feature loss problem a Dense convolutional network or DenseNet was introduced \cite{HuangLW16a}. The DenseNet consists of two major building blocks: dense blocks and transition layers. The dense block solves the vanishing gradient and feature loss problems by reusing features from the previous layers via concatenation, so that all the proceeding layers in the dense block use feature-maps from the previous layers as inputs. This makes possible to maintain low complexity features while also taking advantage of the deep networks ability to probe very complex features of the training data. Such a property makes the DenseNet a great choice when the dataset is somewhat limited and overfitting becomes an issue. The transition layers are then used to reduce the input size. It uses a 1$\times$1 convolutional layer followed by a 2$\times$2 average pooling layer.

\begin{table}[]
\begin{center}
\begin{tabular}{||p{2.2cm} | p{2cm} | p{4.1cm}||} 
\hline
Block & Output size & Layers\\
\hline
Convolution & 134x134x64 & 7x7 conv, stride 2  \\ 
\hline
Pooling & 67x67x64 & 3x3 max pool, stride 2\\
\hline
Dense Block & 67x67x256 & 
$\begin{bmatrix} 
     \text{1x1 conv}\\  
     \text{3x3 conv}\\  
\end{bmatrix}$  x 6\\
\hline
\multirow{2}{2.2cm}{Transition Layer} & 67x67x128 &  1x1 conv \\\cline{2-3}
& 33x33x128 & 2x2 average pooling, stride 2\\
\hline
Dense Block & 33x33x512 & 
$\begin{bmatrix} 
     \text{1x1 conv}\\  
     \text{3x3 conv}\\  
\end{bmatrix}$  x 12 \\
\hline
\multirow{2}{2.2cm}{Transition Layer} & 33x33x256 &  1x1 conv \\\cline{2-3}
& 16x16x256 & 2x2 average pooling, stride 2\\
\hline
Dense Block & 16x16x896 & 
$\begin{bmatrix} 
     \text{1x1 conv}\\  
     \text{3x3 conv}\\  
\end{bmatrix}$  x 20 \\
\hline
\multirow{2}{2.2cm}{Transition Layer} & 16x16x448 &  1x1 conv \\\cline{2-3}
& 8x8x448 & 2x2 average pooling, stride 2\\
\hline
Dense Block & 8x8x1216 & 
$\begin{bmatrix} 
     \text{1x1 conv}\\  
     \text{3x3 conv}\\  
\end{bmatrix}$  x 24 \\
\hline
\multirow{2}{2.2cm}{Output Layer} & 1x1x1216 &  8x8 global average pooling  \\\cline{2-3}
& $N_{out}$ & Fully connected layer with ReLU activation\\
\hline

\end{tabular}
\end{center}
\caption{The structure of the used DenseNet network.}
\label{tab:structure}
\end{table}

In this study we use the DenseNet-BC variant which applies a 1$\times$1 convolutional bottleneck layer before each 3$\times$3 convolution layer in the dense blocks and compression to the transition layer with compression parameter $\theta = 0.5$, which reduces the number of feature-maps by a factor of 2. The growth rate is set to $k = 32$. The DenseNet is originally designed for computer vision tasks and to adapt it to a regression task we change the softmax activation function of the output layer to a linear activation function. The exact structure of the model is shown in Table \ref{tab:structure}, where each convolutional layer contains convolutional layer + batch normalization + ReLU activation. Compared to the original DenseNet model we have changed 3$\times$3 and 7$\times$7 convolution layers with depthwise separable convolution layers, which seems to improve the stability and slightly decrease the validation loss of the model.

\subsection{Implementation}
The DenseNet model is trained using mid-rapidity observables obtained
from the hydrodynamic simulations of heavy-ion collisions computed
in Ref.~\cite{Hirvonen:2022xfv}. The initial energy density profiles for
the hydrodynamic evolution are calculated from the EKRT 
model~\cite{Paatelainen:2013eea, Niemi:2015qia}, where the
event-by-event fluctuations emerge from the random positions of
nucleons inside the colliding nuclei. The computation of the initial 
profiles is very fast and takes a negligible amount of CPU time compared
to the computation of the hydrodynamic evolution and the corresponding
physical observables for each event. It is quite easy to generate
millions of initial conditions corresponding to different collision events.

As an input, the DenseNet model uses  discretized initial
energy density in the transverse-coordinate $(x,y)$ plane
calculated from the EKRT model with a grid size
$269\times269$ and a resolution of 0.07 fm. The DenseNet model is trained to
reproduce a set of final state $p_T$ integrated observables $v_n$,
average transverse momentum $[p_T]$, and charged particle multiplicity
$dN_{ch}/d\eta$ for each event. The input energy density is normalized in such a way
that the training dataset has a mean of zero and a standard deviation of
one. 

The DenseNet model gives then a full event-by-event distribution of these
observables, and it allows us to build a set of measurable
quantities, such as event-averaged $N$-particle flow coefficients
$v_n\{N\}$, normalized symmetric cumulants $NSC(m,n)$, normalized mixed
harmonic cumulants $nMHC(n,m)$ and flow-transverse-momentum correlations
$\rho(v_n^2, [p_T])$. Note that these observables are different moments 
of the full $\mathcal{P}(v_n, [p_T], dN_{ch}/d\eta)$ distribution, 
e.g.\ 2-particle flow coefficient $v_n\{2\}$ is a root-mean-square event-average 
of $v_n$. It is non-trivial that the network can be trained to a 
sufficient accuracy to reproduce these observables, the correlators in particular.
The definitions of all these observables and the details of the EKRT model and 
hydrodynamic computations can be found from Refs.~\cite{Niemi:2015qia, Hirvonen:2022xfv}.


We note that here all the events in a training dataset
use the same parameters for hydrodynamic evolution, meaning that,
currently, the trained neural network cannot predict results from
hydrodynamic simulations that use for example different viscosity
parametrizations.

We train a separate neural network for each of the flow coefficient $v_2, v_3,
v_4, v_5, v_6$, for the average transverse momentum $[p_T]$, and multiplicity
$dN_{ch}/d\eta$ outputs using in total of 20k hydrodynamic events in the
trainining. However, one network can give multiple outputs ($N_{out}$ in Table~\ref{tab:structure})
of the same observable with different $p_T$ integration ranges.
This is necessary since different measurement use different $p_T$ ranges
when measuring the observables. 

The training events are distributed
evenly (5k events each) between 200 GeV Au+Au, 2.76 TeV Pb+Pb, 5.023 TeV
Pb+Pb and 5.44 TeV Xe+Xe collision systems. The outputs of different
neural networks are normalized with a constant such that the typical
value of a given output observable is $\mathrm{O}(1)$. This makes
possible to set the same learning rate for different observables without
affecting the quality of the training too much. The exception to this is
the charged particle multiplicity network for which the output is not
normalized because it uses a different loss function than the other
networks. The training data is heavily
augmented by applying random rotations (rotation angle from 0 to
$2\pi$), flips and translations (shifts from -0.92 fm to 0.92 fm in both
x and y directions) to the input during the training.

All the network models above are trained using the Adam optimizer~\cite{Kingma14} for 120
epochs with a batch size of 64. Using larger batch sizes made the training
phase faster, but at the same time significantly decreased the accuracy of
the networks. The learning rate is initially set to 0.001, except in the
case of the charged particle multiplicity where the initial learning rate
is 0.01, and it is divided by a factor of ten at epochs 75 and 110. Even
though the use of a decaying learning rate is not completely necessary
because of the adaptive nature of the Adam optimizer, we noticed that
adding a learning rate decay made the training faster without
sacrificing accuracy. Additionally the batch normalization momentum is
set to 0.1. As a regularization method we tried both the dropout and L2
regularization, but they did not give any improvements for the validation
accuracy or made it worse. This is most likely due to a heavy data
augmentation which in itself acts as an efficient regularization method.

For all observables except charged particle multiplicity, we use a mean squared error (MSE) loss function which is defined as
\begin{equation}
    \mathrm{Loss(MSE)} = \frac{1}{N} \sum_{i} (y_{\rm{i, true}} - y_{\rm{i, pred}})^2,
\end{equation}
where the sum is over all events in the training batch, $N$ is the number of events in a training batch and $y_{\rm{i, true}}$ and $y_{\rm{i, pred}}$ are the true and predicted values of an observable respectively. For the charged particle multiplicity we use a mean squared logarithmic error (MSLE) loss function,
\begin{equation}
    \mathrm{Loss(MSLE)} = \frac{1}{N} \sum_{i} (\ln(y_{\rm{i, true}}+1) - \ln(y_{\rm{i, pred}}+1))^2.
\end{equation}

The training is done using the Nvidia Tesla V100 GPU, which has 32 GB of VRAM and 640 tensor cores. The training time for one network is ca.\ 80 min.
The neural network code is written in Python and it is implemeted using the Keras Deep Learning API v2.10.0 \cite{Chollet} together with the Tensorflow v2.10.0 library \cite{Abadi16}.

\begin{figure*}
\includegraphics[width=\textwidth]{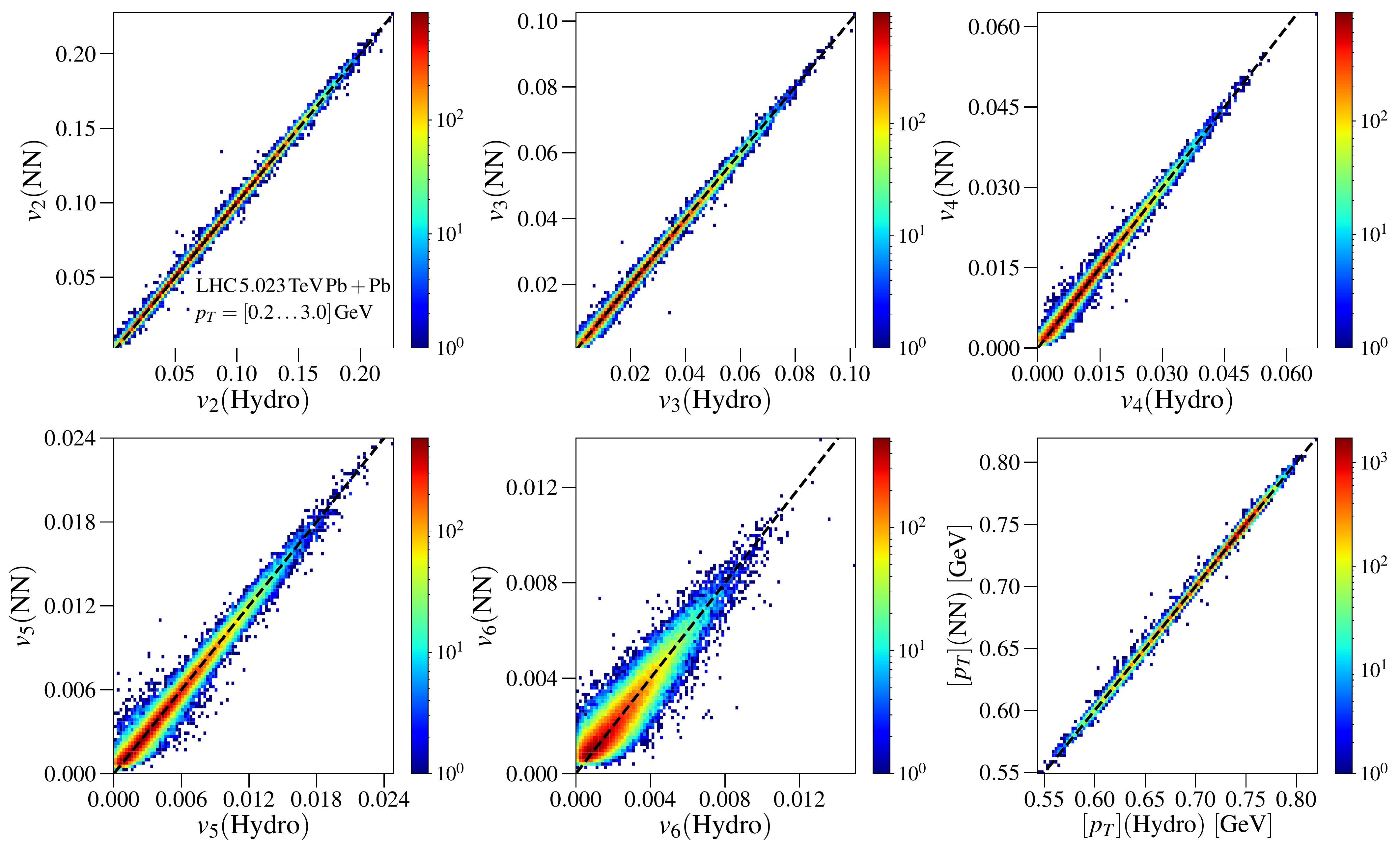}
\caption{(Color online) The event-by-event neural network (NN) predictions versus the results from the hydrodynamic simulations for the validation events in the 0-80\% centrality range.}
\label{fig:NN_error}
\end{figure*}

\begin{figure*}
\includegraphics[width=\textwidth]{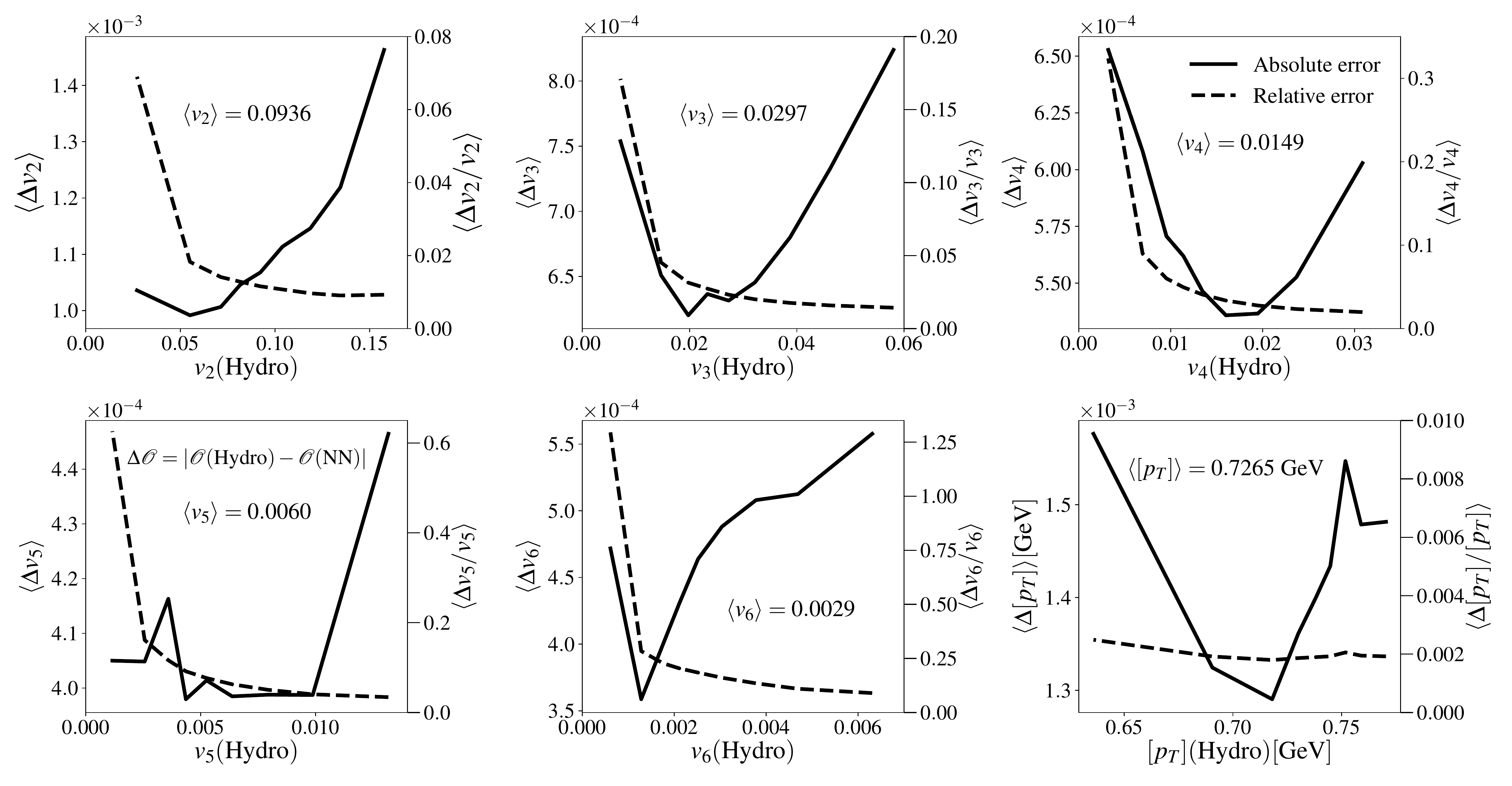}
\caption{(Color online) The mean absolute and relative errors between the neural network predictions and the results from the hydrodynamic simulations for the validation events in the 0-80\% centrality range.}
\label{fig:NN_MAE}
\end{figure*}

\begin{figure*}
\includegraphics[width=\textwidth]{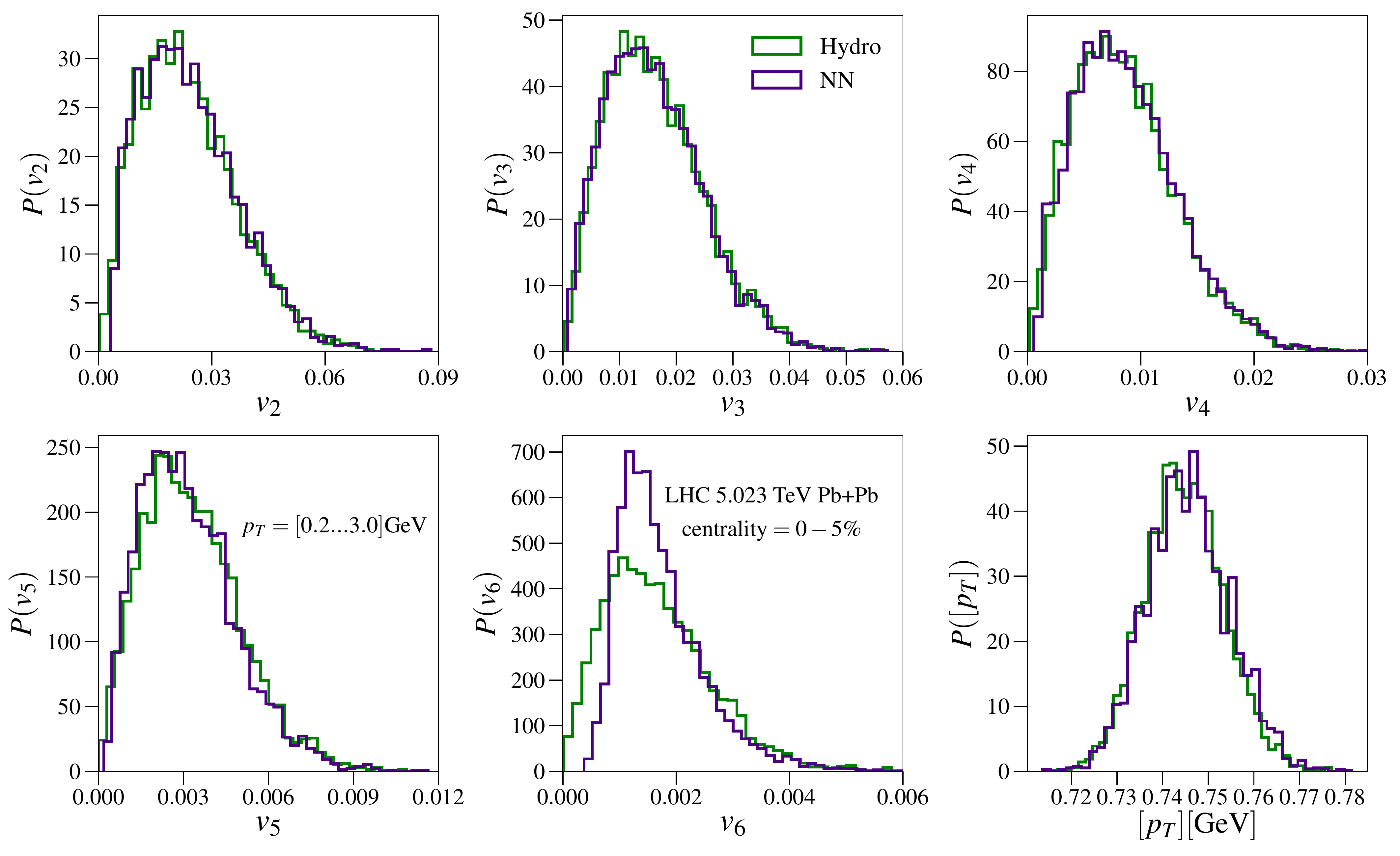}
\caption{(Color online) The distributions of flow observables from the neural network predictions and hydrodynamic simulations for the validation events in the 0-5\% centrality range.}
\label{fig:NN_dist}
\end{figure*}

\section{Validation}
\label{sec:validation}
After the training, the accuracy of the neural network needs to be tested with an independent validation dataset. Here we only focus on results for a 5.023 TeV Pb+Pb collision system, but the performance of the neural network is similar for other systems as well.
The testing is done by generating 90k initial energy density profiles and comparing neural network predictions for different observables against those obtained from hydrodynamic simulations. We remind that only 5k 5.023 TeV events were used in the training of the network.

\begin{figure*}
\includegraphics[width=\textwidth]{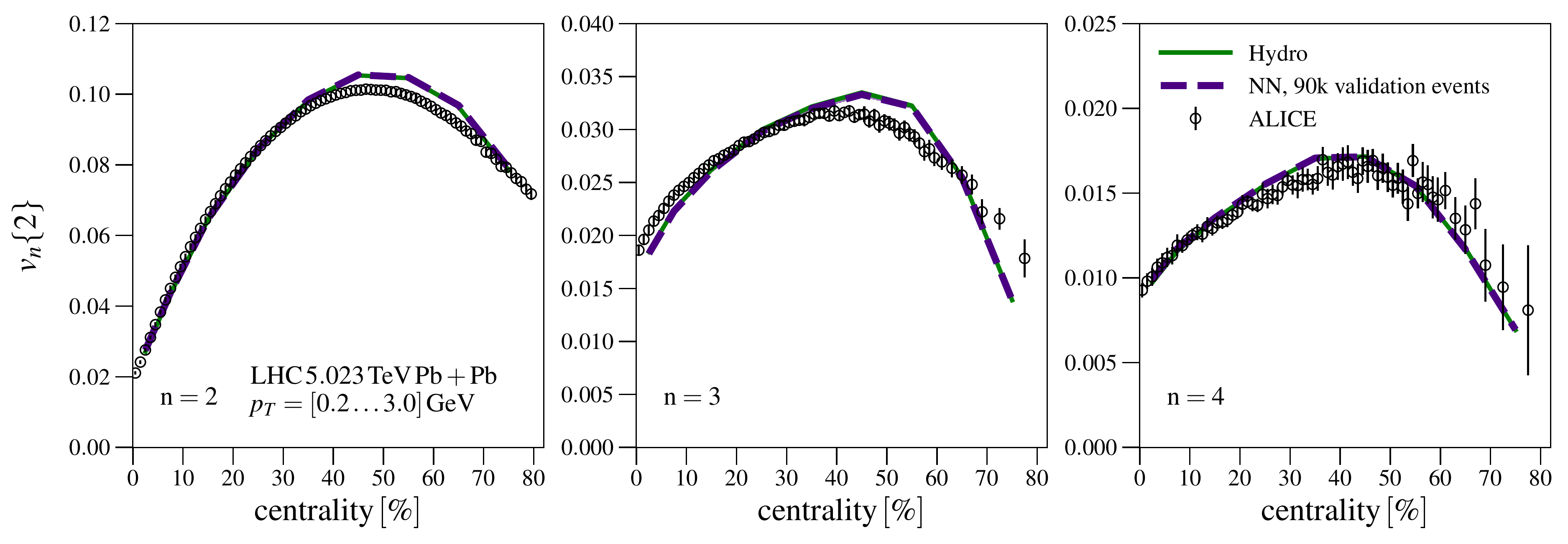}
\includegraphics[width=\textwidth]{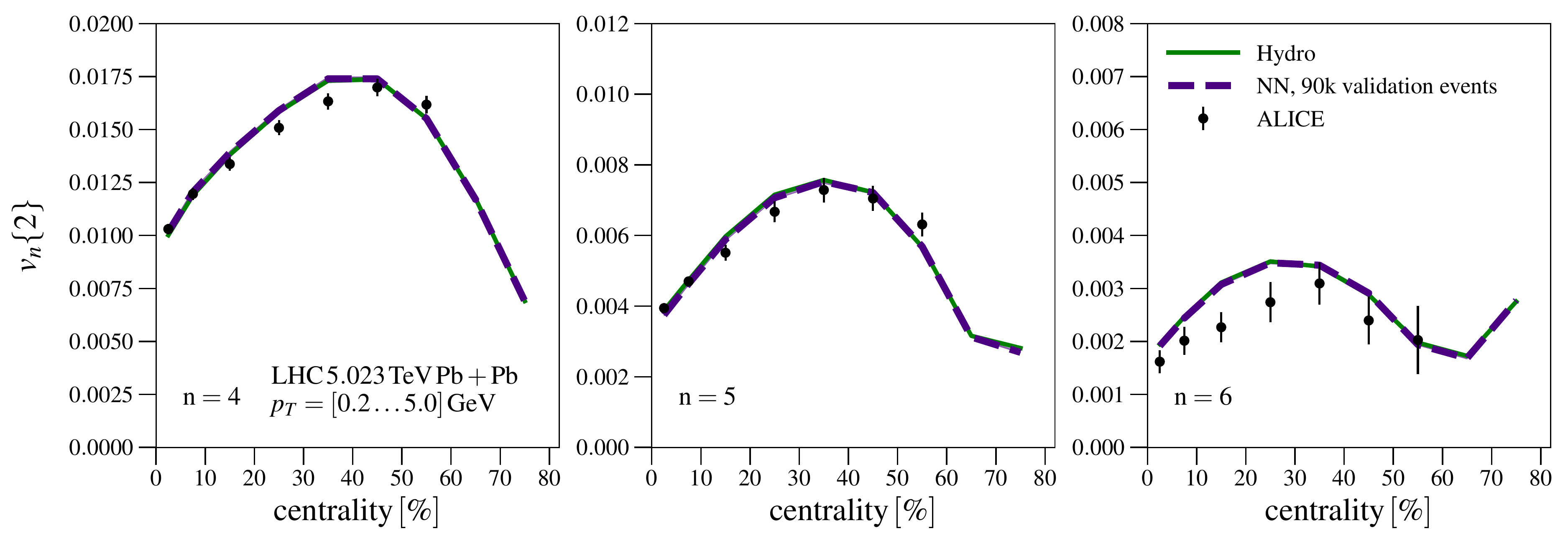}
\caption{(Color online) The comparison of the flow coefficients $v_n\{2\}$ between the neural network predictions and hydrodynamic computations. The experimental data are from the ALICE Collaboration \cite{ALICE:2018rtz,ALICE:2020sup}.}
\label{fig:vnitegrated_val}
\end{figure*}

In Fig.~\ref{fig:NN_error}, we show a 2d histogram comparing the neural network predictions against hydrodynamic computations event-by-event for the flow coefficients $v_n (n = 2,3,4,5,6)$ and average transverse-momenta $[p_T]$.  The color bar indicates the number of events in each histogram bin and the dashed black line indicates where hydrodynamic computations and neural network predictions match exactly. Because the observables we are interested in are inside the 0-80\% centrality range, we only show events from this centrality range in the histogram.

For $v_2$ we see an excellent agreement between the neural network and hydrodynamic results. The accuracy of the network starts to slowly decrease when moving towards higher-order flow coefficients and in the cases of $v_5$ and $v_6$ we already start to see clear deviations from the hydrodynamic results. This behavior is expected since the lower-order flow coefficients and initial-state eccentricities have quite linear dependence and they are not as sensitive to nonlinear effects arising from hydrodynamic evolution as higher-order flow coefficients. For the average transverse momentum the neural network seems to predict the hydrodynamic results very accurately. However, one needs to note that event-by-event fluctuations of $[p_T]$ are very small compared to the absolute value of $[p_T]$. This means that relatively small errors are not necessarily a guarantee of that the network can correctly predict correlations involving $[p_T]$.

To complement the information in Fig.~\ref{fig:NN_error} and to give more quantitative estimates of errors, we show the mean absolute and the relative errors for different observables in Fig.~\ref{fig:NN_MAE}. Here we can confirm that the relative error is indeed increasing when increasing the order of the flow coefficients. The errors are not very sensitive to the value of an observable but typically the absolute errors are smallest close to the average value of the observable. We can also notice that the relative error is the largest when the value of an observable is small. The small values of flow coefficients usually correspond to the most central or the most peripheral collisions.

To see where the growing relative errors at the smallest values of $v_n$ start to play a role, we compare distributions of flow observables between the neural network predictions and hydrodynamic computations in the most central collisions. The results are shown in Fig.~\ref{fig:NN_dist}, where we can see that the distributions are nearly identical except for the flow coefficient $v_6$. In this case the distribution given by the neural network prediction is narrower than the one obtained from the hydrodynamic computation while the location of the peak value is very similar in both cases. This indicates that the neural network might be able to reproduce the average values of $v_6$ quite well but it cannot be guaranteed to reliably predict the correlations involving $v_6$.

\begin{figure*}
\includegraphics[width=\textwidth]{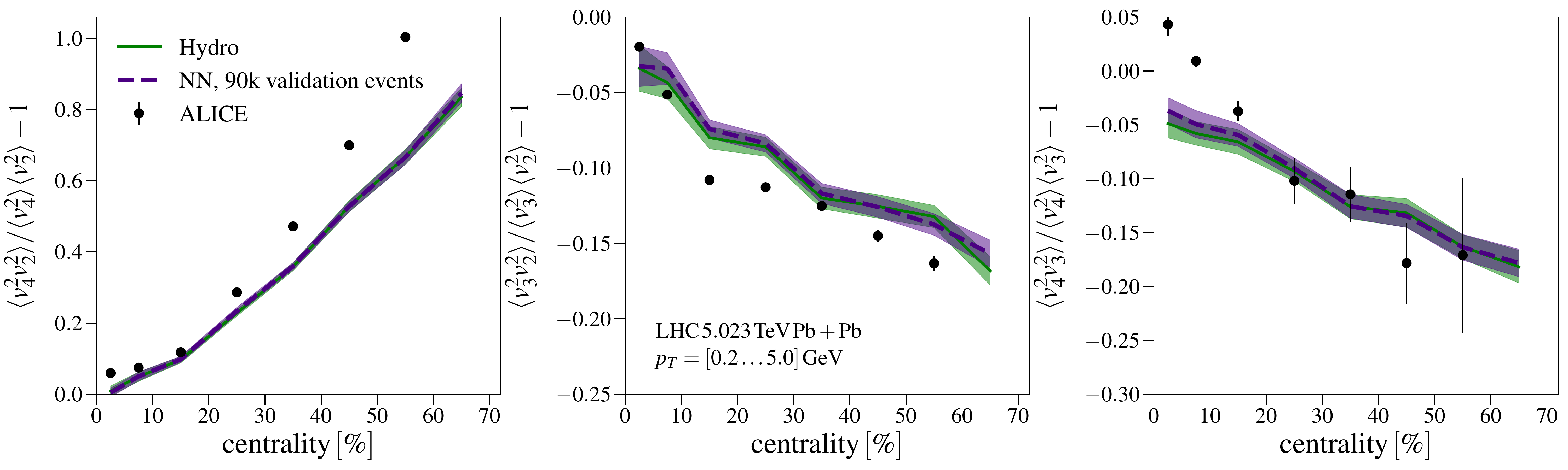}
\caption{(Color online) The comparison of normalized symmetric cumulants between the neural network predictions and hydrodynamic computations. The experimental data are from the ALICE Collaboration \cite{ALICE:2021adw}.}
\label{fig:nsc_val}
\end{figure*}
\begin{figure*}
\includegraphics[width=\textwidth]{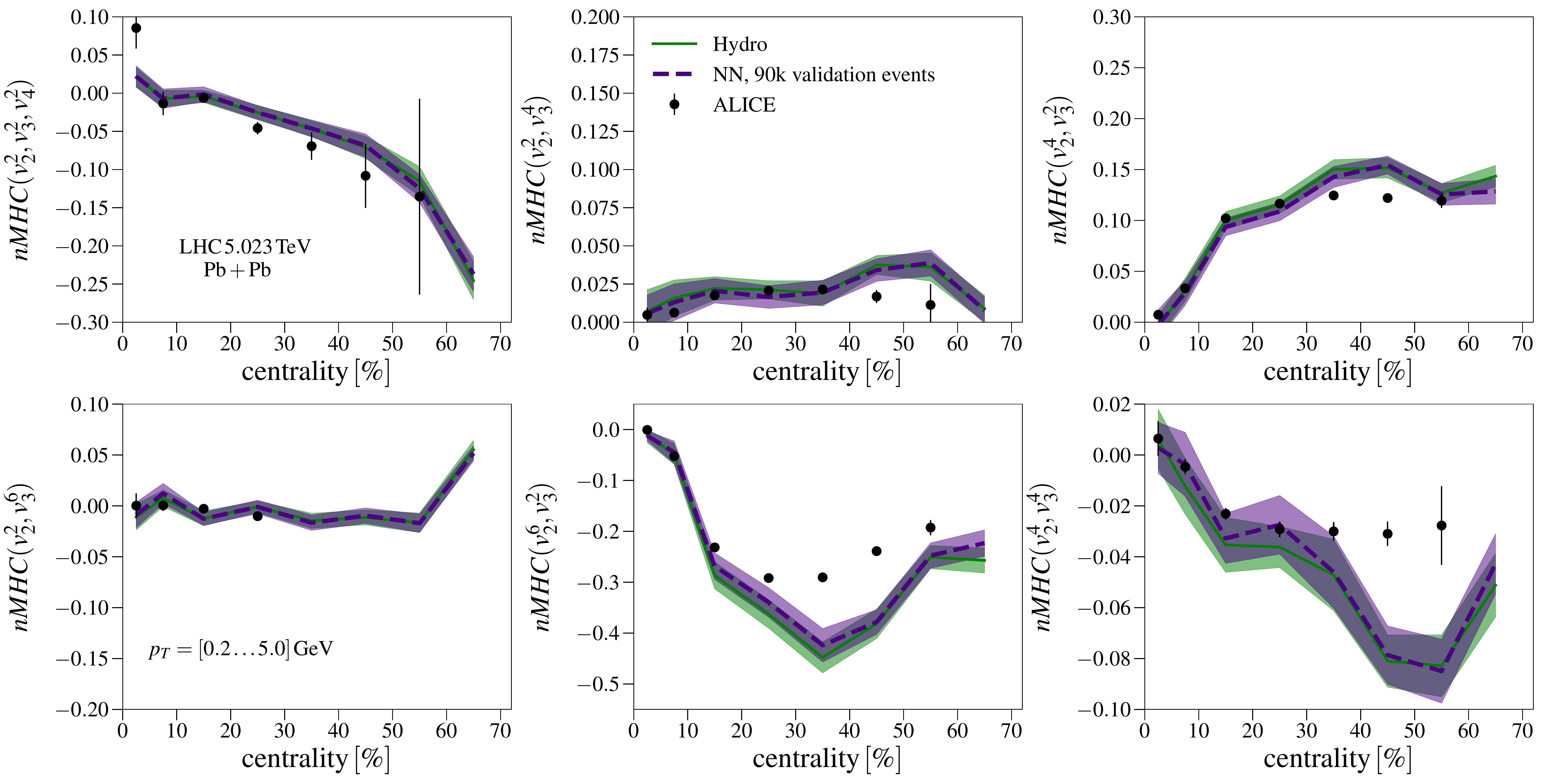}
\caption{(Color online) The comparison of normalized mixed harmonic cumulants between the neural network predictions and hydrodynamic computations. The experimental data are from the ALICE Collaboration \cite{ALICE:2021adw}.}
\label{fig:nmhc_val}
\end{figure*}
\begin{figure*}
\includegraphics[width=\textwidth]{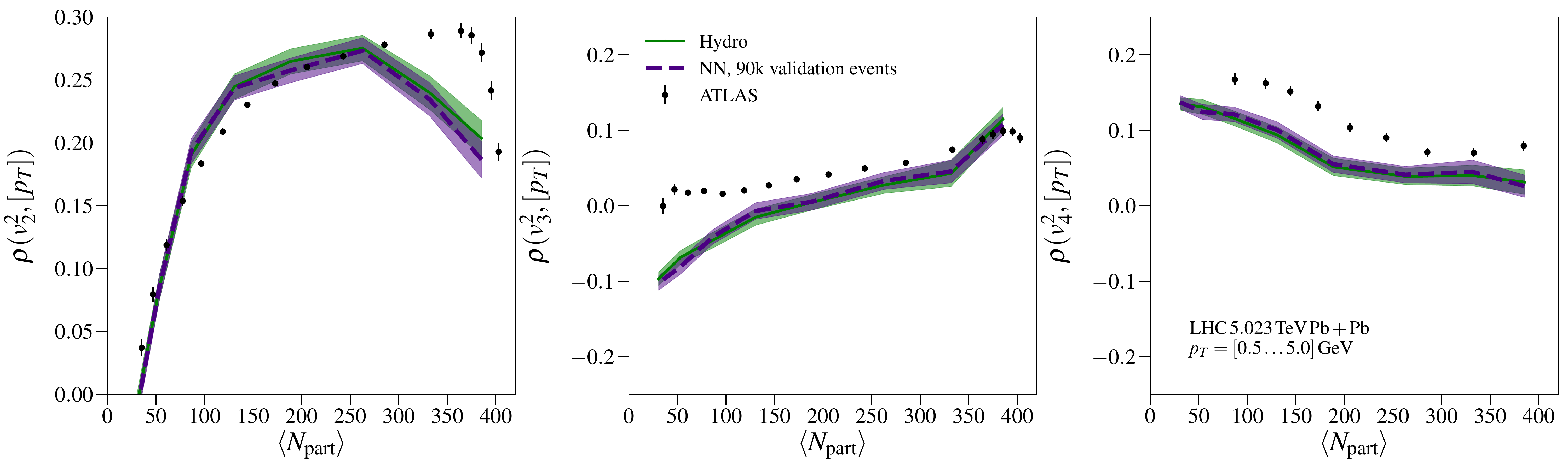}
\caption{(Color online) The comparison of flow-transverse-momentum correlations between the neural network predictions and hydrodynamic computations. The experimental data are from the ATLAS Collaboration \cite{ATLAS:2019pvn}.}
\label{fig:rhov2pt_val}
\end{figure*}

Comparing the neural network and hydrodynamic results event-by-event gives information about the accuracy of the network, but the measurements average over a large number of events in centrality bins. Consequently, it is crucial to test the performance of the network in these cases as well. To get a comprehensive view of the network's ability we check its performance for 2-particle flow coefficients $v_n\{2\}$, normalized symmetric cumulants $NSC(m,n)$, normalized mixed harmonic cumulants $nMHC(n,m)$ and flow-transverse-momentum correlations $\rho(v_n^2, [p_T])$. 

The flow coefficients $v_n\{2\}$ are shown in Fig.~\ref{fig:vnitegrated_val} as a function of centrality. We can see that the neural network results seem to match the hydrodynamic results nearly exactly. This is true even in the cases of $v_5$ and $v_6$ where the event-by-event accuracy of network was not as good.

Much more challenging quantities to predict are the different correlations between the flow coefficients. In Fig.~\ref{fig:nsc_val} we show the centrality dependence of the normalized symmetric cumulants $NSC(m,n)$. The statistical errors are estimated via jackknife resampling as in Ref.~\cite{Hirvonen:2022xfv}. The normalized symmetric cumulants are four-particle correlations between two flow harmonics and thus are more sensitive to event-by-event fluctuations than the flow coefficients $v_n\{2\}$. This makes it more challenging to predict them using the neural network. Nevertheless, in the case of $NSC(4,2)$ we get an almost exact agreement between the neural network and the hydrodynamic results. For $NSC(3,2)$ and $NSC(4,3)$ there are some visible differences between the two, but deviations are still quite small compared to the statistical errors.

The normalized mixed harmonic cumulants $nMHC(n,m)$, which are 6- or 8-particle correlations, are shown in Fig.~\ref{fig:nmhc_val}. The agreement between the neural network predictions and the hydrodynamic computation is again good, even in the cases where the correlation is very weak. Finally, in Fig.~\ref{fig:rhov2pt_val}, we show the flow-transverse-momentum correlations $\rho(v_n^2, [p_T])$ as a function of the number of participant nucleons. In this observable the biggest challenge for the neural network is not the accuracy of the flow coefficients as one might naively expect, but instead the accuracy of the mean transverse momentum. This is due to the fact that the correlation is very sensitive to the mean transverse momentum fluctuations and, as discussed earlier, catching these fluctuations requires a very good precision from the neural network. Nevertheless, as can be seen from Fig.~\ref{fig:rhov2pt_val}, the neural network predictions agree well with the hydrodynamic results.

\begin{figure*}
\includegraphics[width=\textwidth]{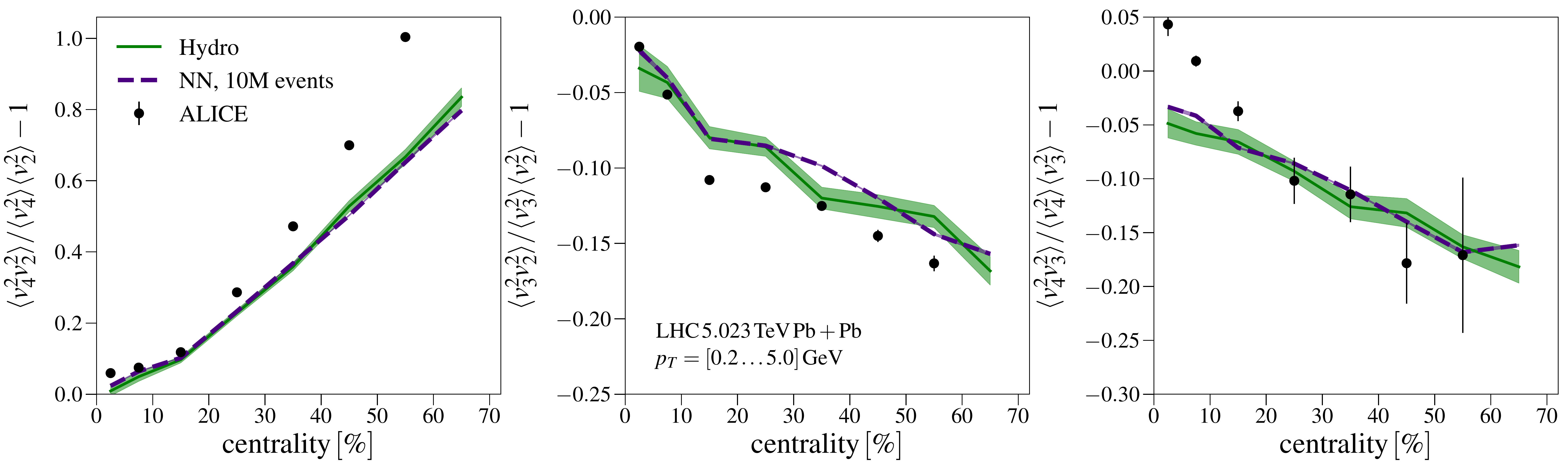}
\caption{(Color online) The neural network prediction of normalized symmetric cumulants with 10M collision events compared with the hydrodynamic results from 90k collision events. The experimental data are from the ALICE Collaboration \cite{ALICE:2021adw}.}
\label{fig:nsc_highstat}
\end{figure*}
\begin{figure*}
\includegraphics[width=\textwidth]{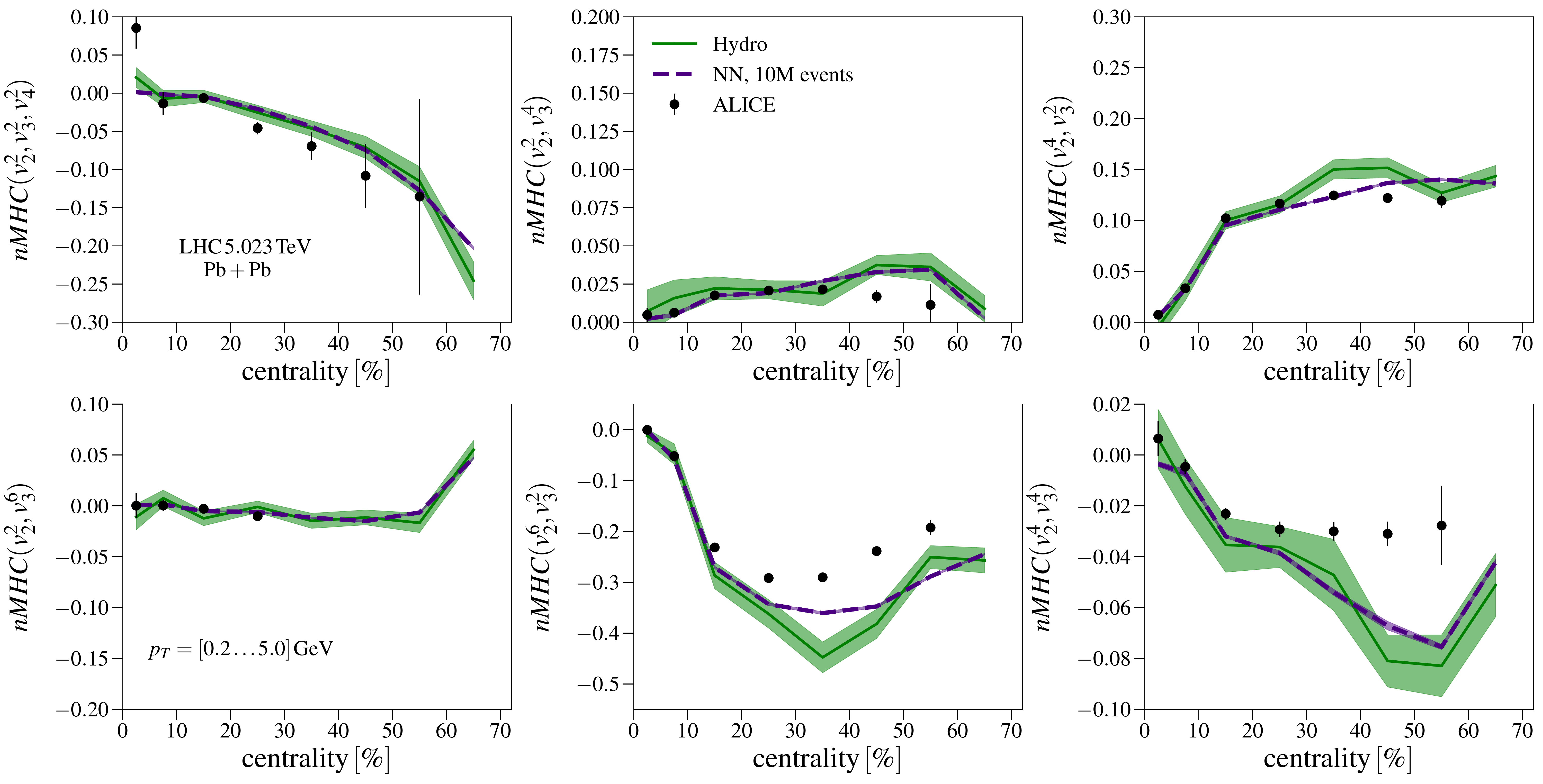}
\caption{(Color online) The neural network prediction of normalized mixed harmonic cumulants with 10M collision events compared with the hydrodynamic results from 90k collision events. The experimental data are from the ALICE Collaboration \cite{ALICE:2021adw}.}
\label{fig:nmhc_highstat}
\end{figure*}
\begin{figure*}
\includegraphics[width=\textwidth]{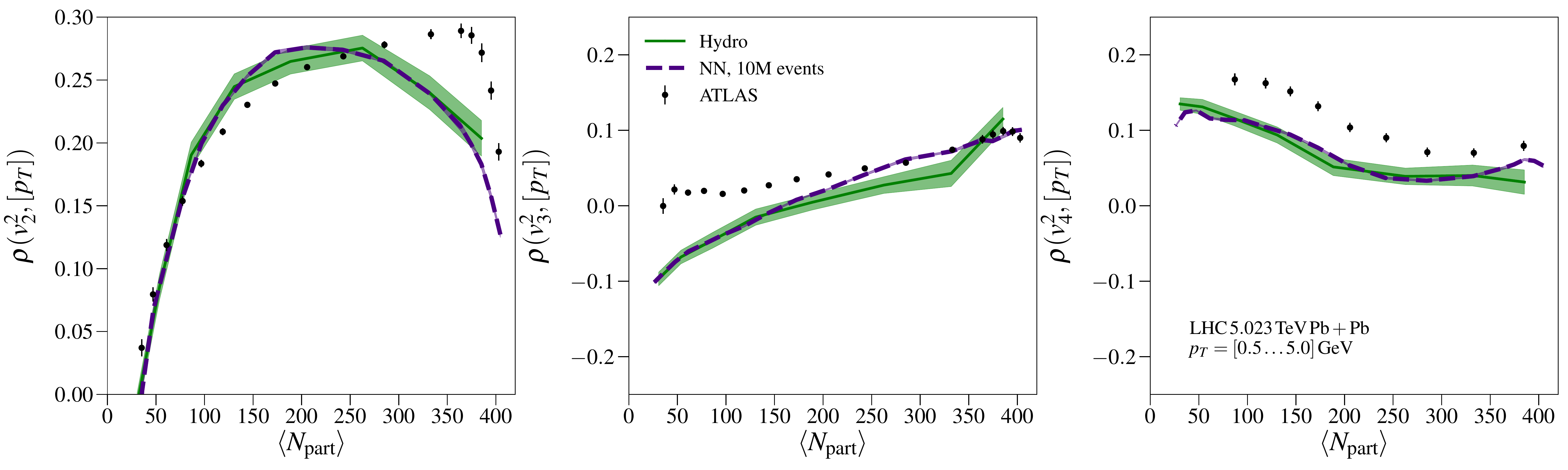}
\caption{(Color online) The neural network prediction of flow-transverse-momentum correlations with 10M collision events compared with the hydrodynamic results from 90k collision events. The experimental data are from the ATLAS Collaboration \cite{ATLAS:2019pvn}.}
\label{fig:rhov2pt_highstat}
\end{figure*}

\section{High-statistics predictions}
\label{sec:results}
Now that the accuracy of the neural network has been established, we can use it to estimate what happens to the above correlations at a high-statistics limit. To do so we generate 10M events using the neural network, which takes around 20 hours with the GPU. This is a very substantial difference compared to doing full hydrodynamic simulations using CPU, which would take $\sim 5$M CPU hours.

The effect of increased statistics for the normalized symmetric cumulants can be seen in Fig~\ref{fig:nsc_highstat}. In the case of $NSC(4,2)$ we see slight deviations in the most central and peripheral collisions, but the centrality dependence is very similar to the lower statistics hydrodynamic results. This is not surprising since the statistical errors are already relatively small with 90k events. The situation is quite different for $NSC(3,2)$ where the statistical errors are of considerable size with 90k events. Here we see that with 10M events the statistical fluctuations are negligible, revealing the true centrality dependence from the model, and it now gives a very similar shape as the ALICE measurements, even though the neural network prediction (i.e.\ the underlying hydrodynamic simulation with which the network was trained) underestimates amount of anti-correlation. For $NSC(4,3)$ we also see some deviations from the lower-statistics hydrodynamic result in the most central and peripheral collisions. We note that in the most central collisions we see somewhat similar difference between the neural network and the hydrodynamic result also in the validation dataset, which might indicate that this difference can be a systematic error caused by inaccuracy of the neural network.

In principle, the normalized mixed harmonic cumulants in Fig.~\ref{fig:nmhc_highstat} should be even more sensitive to the increased event number, since correlations are usually weaker than in the case of the normalized symmetric cumulants. For $nMHC(v_2^2, v_3^4)$ the neural network prediction with 10M events is inside the statistical errors of the hydrodynamic results, but in the central collisions the increased number of events reveals very different kind of centrality dependence which seems to agree well with the ALICE measurements. In the cases of $nMHC(v_2^4, v_3^2)$ and $nMHC(v_2^6, v_3^2)$ we see statistically significant differences between the hydrodynamic results and 10M event predictions, which signals that the jackknife resampling can sometimes significantly underestimate the statistical errors. For $nMHC(v_2^2, v_3^6)$ we see that increasing the number of events from 90k to 10M removes the sharp changes between the correlation and anti-correlation and the high statistic result is nearly zero except in the most peripheral collisions. This is again in line with the ALICE measurements.

The flow-transverse-momentum correlations for the 10M neural network prediction are shown in Fig.~\ref{fig:rhov2pt_highstat}. The increased statistics makes it now possible to use exactly the same centrality bins as the ATLAS measurements without completely ruining the accuracy. For $\rho(v_2^2, [p_T])$ the 10M event result differs substantially from the 90k event hydrodynamic result only in the most central collisions. This effect is mostly a combination of different centrality binning and the fact that correlation decreases very quickly when moving from 375 to 400 participants. The effect of statistics can be better seen in the case of $\rho(v_3^2, [p_T])$, where in the central collisions the 10M event result has different dependence on participant number than the 90k event hydrodynamic result. In this region the 10M event neural network result also agrees better with the ALICE measurements.

\section{Conclusions and summary}
\label{sec:conclusions}
We have trained a deep convolutional neural network to predict a variety of flow observables from the initial state energy density profiles. The training was done using 20k training events from 200 GeV Au+Au, 2.76 TeV Pb+Pb, 5.023 TeV Pb+Pb and 5.44 TeV Xe+Xe collision systems, with 5k events for each collision system. The training data was computed using viscous relativistic hydrodynamics with initial conditions from the EKRT model, and using the model and viscosity parameters from Ref.~\cite{Hirvonen:2022xfv}.

The accuracy of the network was tested against the results from hydrodynamic simulations for 2-particle flow coefficients $v_n\{2\}$, normalized symmetric cumulants $NSC(m,n)$, normalized mixed harmonic cumulants $nMHC$ and flow-transverse-momentum correlations $\rho(v_n^2, [p_T])$. We emphasize that this is a non-trival test for the accuracy of the network, especially with the correlators. The validation tests used in total of 90k events for each collision system, independent of the training data, and in all of the cases the neural network was able to predict hydrodynamic results quite reliably. This is already a significant improvement in terms of computational time, as only 5k events were used per collision system to train the network.

The neural network was then used to predict the same flow observables but this time with 10M generated events. This took around 20 GPU hours of computing time which is many orders of magnitude faster than doing the same number of hydrodynamic simulations using CPU. The increased number of events made statistical errors negligibly small and allowed us to estimate the observables with a higher precision. In many cases the 10M event neural network prediction differed from the 90k event hydrodynamic computations by a quite large margin emphasizing the importance of a large event statistics when comparing simulations with the measurements.

As there are still considerable uncertainties in determining QCD matter properties from the experimental data, it is important to be able to use as many measurements as possible to constrain the properties. In particular, the current measurements at the LHC give a wealth of different flow correlations with tight error bars that provide independent information about the matter properties. Many of the measured correlators are rather weak, and can require millions of computed hydrodynamic events in order to get similar statistical errors as in the experiments. In order to use these quantities as a constraint to the QCD properties, it then necessary to have a computationally efficient way to generate such large set of events, and this is exactly what the neural network presented here can do.

Currently the neural network can predict flow observables for different initial energy density profiles, but the predictions always describe a hydrodynamic evolution that is identical to the one used in the training dataset, i.e.\ it is not possible to change the viscosity parametrization after the network has been trained. The network should next be constructed to be more versatile and take viscosity parameters as additional inputs, making the neural network even more efficient tool in global analysis. This is left as future work.

\acknowledgments
We acknowledge the financial support from the Jenny and Antti Wihuri Foundation, and the Academy of Finland project 330448 (K.J.E.). This research was funded as a part of the Center of Excellence in Quark Matter of the Academy of Finland (project 346325). This research is part of the European Research Council project ERC-2018-ADG-835105 YoctoLHC. The Finnish IT Center for Science (CSC) is acknowledged for the computing time through the Project jyy2580.

\end{document}